\documentclass[11pt]{article}

\usepackage{graphicx}
\usepackage{subfigure}
\usepackage{amsfonts,amssymb,amsmath}
\usepackage{vmargin}
\usepackage{cite}

\numberwithin{equation}{section}

\newcommand{\om}{\ensuremath{\omega}}

\newcommand{\M}{\ensuremath{{\cal M}}}
\newcommand{\N}{\ensuremath{{\cal N}}}

\newcommand{\ra}{\ensuremath{\rightarrow}}

\newcommand{\half}{\ensuremath{\frac{1}{2}}}

\newcommand{\be}{\begin{equation}}
\newcommand{\ee}{\end{equation}}

\newcommand{\ba}{\begin{eqnarray}}
\newcommand{\ea}{\end{eqnarray}}

\newcommand{\ns}{\normalsize}

\newcommand{\gsim}{\raise.3ex\hbox{$>$\kern-.75em\lower1ex\hbox{$\sim$}}}
\newcommand{\lsim}{\raise.3ex\hbox{$<$\kern-.75em\lower1ex\hbox{$\sim$}}}

\newcommand{\nn}{\nonumber}

\newcommand{\w}{\wedge}

\newcommand{\im}{\mathrm{Im\;}}
\newcommand{\re}{\mathrm{Re\;}}

\newcommand{\href}[1]{\underline{#1}}

\begin{document}

\begin{titlepage}
 
\title{
   {\Large\bf Model building with intersecting D6-branes on smooth Calabi-Yau manifolds}
      \\[0.5cm]
}
      
\author{
{\ns\large 
  Eran Palti \footnote{email: palti@thphys.ox.ac.uk}
}
\\[0.5cm]
   {\it\ns Rudolf Peierls centre for Theoretical Physics, University of Oxford}\\
   {\it\ns Keble Road, Oxford, UK. } 
}
\date{}

\maketitle

\begin{abstract}\noindent
We study intersecting D6-branes in Calabi-Yau manifolds that are smooth hypersurfaces in weighted projective spaces. We develop the techniques for calculating intersection numbers between special Lagrangian sub-manifolds defined as fixed loci of anti-holomorphic involutions. We present global Pati-Salam and MSSM-like models that are supersymmetric up to a decoupled hidden sector.       
\end{abstract}
 
\thispagestyle{empty}
 
\end{titlepage}

\pagestyle{plain}
\setcounter{page}{1}
\newcounter{bean}
\baselineskip14pt

\tableofcontents

\newpage
\section{Introduction}

String theory model building is an important part of string phenomenology. One sector where realistic chiral gauge theories can be constructed is intersecting D6-branes in type IIA string theory \cite{Berkooz:1996km}. For a review on the subject see \cite{Blumenhagen:2006ci}. The vast majority of work so far has studied intersecting branes on orbifolds of tori. These are in many ways the simplest Calabi-Yau (CY) manifolds but form only a very small group out of the known set of CY manifolds.\footnote{Realistic chiral models within the heterotic string have been constructed on smooth CYs for a long time, see \cite{Anderson:2008uw} for state-of-the-art. Within type IIB see \cite{Blumenhagen:2008zz} for state-of-the-art.}
In this paper we develop further type IIA model building with intersecting D6-branes on a larger group of CY manifolds that is composed of smooth hypersurfaces in possibly weighted projective spaces. These are spanned by the complete intersection CYs (CICYs) (of which there are 7890 \cite{Candelas:1987kf}) and by one Kahler parameter hypersurfaces in weighted projective spaces (of which there are 3 \cite{Candelas:1989hd}).

In order to preserve supersymmetry the space-filling D6-branes must wrap special Lagrangian sub-manifolds of the CY \cite{Becker:1995kb}. Therefore model building amounts to identifying a set of special Lagrangian sub-manifolds and calculating their intersections. 
In \cite{Brunner:1999jq} a systematic construction of special Lagrangians and their intersections was performed for the quintic . The model building application for that case were studied in \cite{Blumenhagen:2002wn,Blumenhagen:2002vp} where it was shown that a chiral and gauge spectrum of exactly the standard model can be obtained. This is a promising result given only a single case has been analyzed. The major problem with the quintic constructions however is that it was not possible to have a chiral spectrum and preserve supersymmetry simultaneously since any special Lagrangian sub-manifolds that shared a calibration did not intersect. In this paper we study intersecting brane model building on other CY manifolds and show that these do have sets of special Lagrangian sub-manifolds that share a calibration and have net intersection number. This opens up the possibility of supersymmetric model building which forms our aim. Since just the CICYs number in the thousands a classification of realistic models on all the manifolds is beyond the scope of this work. Rather we set out to develop the tools needed to study any chosen manifold and only study explicit models for the most promising cases. 

An important distinction within intersecting brane model building (in type IIA or IIB) is between local and global models. Local models are studied within a local non-compact geometry without an explicit embedding into a compact manifold. Within type IIA this has been studied in \cite{Uranga:2002pg}.\footnote{In type IIB/F-theory of course the subject of local models is much more developed. See \cite{Beasley:2008dc} for state-of-the-art.} Global models are studied on the full compact geometry. Both have advantages and disadvantages. Within this classification the models in this paper are global. Therefore they should enjoy the advantages of being closer to global issues such as tadpole cancellation, supersymmetry breaking and moduli stabilisation. However it turns out that for the explicit cases studied the special Lagrangian set does not span the full homology of the manifold. This  means that practically these models behave much like local models: there is always a sector which needs to be added to ensure tadpoles are satisfied and $U(1)$s are massless which is not explicitly known. This then brings with it some of the disadvantages of local models such as lack of control over supersymmetry breaking in the hidden sector. It is important to state that this `locality' is not a property of the methodology but rather of the geometry of the particular cases studied. Different CYs can avoid this issue (indeed the quintic does avoid this problem but is faced with the supersymmetry problem). More details on this issue are given in the main body of the paper but it is important enough to state from the offset. 

The layout of the paper is as follows. In section \ref{sec:fslags} we show how to identify special Lagrangian sub-manifolds using anti-holomorphic involution symmetries of the CY. In section 
\ref{sec:intsplag} we show how to calculate their intersections. In section \ref{sec:intbran} we discuss how the geometry of the previous sections applies to intersecting brane model building. In section \ref{sec:modbuild} we present some two example models that are a supersymmetric Pati-Salam model and an MSSM-like model (with chiral exotics). In appendices \ref{sec:curandsur} and \ref{sec:intexam} we calculate some intersection numbers and present an example calculation. In appendix \ref{sec:someothermodels} we present some more example models.

\section{Identifying special Lagrangian submanifolds}
\label{sec:fslags}

In this section we study a way special Lagrangian submanifolds can be identified within CY manifolds. We use this method in all our constructions. The CY manifolds that we consider are given by  a number of holomorphic polynomial equations within complex projective spaces. We only study smooth manifolds. Many smooth CYs can be constructed by blowing up singular hypersurfaces within weighted projective spaces. We do not include these in our analysis and only consider hypersurfaces that are smooth. This drastically cuts down the number of candidate manifolds. We leave a study of the `singular' manifolds for future work.  

The manifolds that we consider are the set of Complete Intersection CYs (CICYs) that are hypersurfaces within a product of ordinary projective spaces and the three smooth cases of hypersurfaces within weighted projective spaces. There are 7890 CICYs which were constructed in \cite{Candelas:1987kf} and they are all smooth. CYs as blow-ups of hypersurfaces in weighted projective spaces were constructed in \cite{Candelas:1989hd}. Weighted projective spaces are always singular (with the exception of the trivial case of all weights equal). However the dimension of the singularity is given by the number of weights that share common factors. Therefore weighted projective spaces with co-prime weights only have point singularities. Of the 7555 cases constructed in \cite{Candelas:1989hd} only 120 have co-prime weights. Of these the hypersurface will miss these singularities, and so the CY will be smooth, if and only if the polynomial is of Fermat type which means that each weight must divide the sum of the weights. This leaves three (plus the quintic) cases which we list in table \ref{wcyexa}. Note that we have picked a point in moduli space where all the monomials vanish and only the defining polynomial remains.\footnote{This is a stronger condition then is actually necsessary. Indeed any monomial which respects the symmetry of the anti-holomprhic involutions (see section \ref{sec:slagfromiso}) responsible for the special Lagrangians that the branes in a particular model wrap can have an arbitrary coefficient. For example in supersymmetric models (see section \ref{sec:countslags}) the coefficient of the fundamental monomial is arbitrary.}
The question of whether this is a dynamically preferred point in moduli space is beyond the scope of this paper.

Throughout this section we denote the vanishing polynomials that define the CY hypersurface within the ambient space as $P_A(z)$ where $z$ are the coordinates on the ambient space and the index $A$ runs over the number of polynomials. 

Throughout this section we keep the formulas as general as possible. In appendix \ref{sec:intexam} we go through an explicit example computation of an intersection between two special Lagrangians in weighted projective space which serves as a clarifying example for the general techniques outlined in this section.

\subsection{Special Lagrangians from isometric anti-holomorphic involutions}
\label{sec:slagfromiso}

For a mathematical review of special Lagrangian manifolds see \cite{Joyce:2001nm}. If we have a CY manifold $\M$ with a metric $g$, and Kahler form $J$ and a holomorphic three-form $\Omega$, then a special Lagrangian sub-manifold is a three-dimensional sub-manifold $\Pi$ such that when $J$ and $\Omega$ are pulled back to it they satisfy
\be
\left. \im{\left(e^{i\frac{\theta_{\Pi}}{2}}\Omega\right)}\right|_{\Pi} = 0 \;,\; \left.J\right|_{\Pi} = 0 \;, \label{slagdef}
\ee           
for some angle $\theta_{\Pi}$ that is associated with $\Pi$. 
Special Lagrangian manifolds are volume minimising in their homology class and are calibrated so that their volume form $\epsilon_{\Pi}$ is given by
\be
\epsilon_{\Pi} = \left. \re{\left(e^{i\frac{\theta_{\Pi}}{2}}\Omega\right)}\right|_{\Pi}. \label{slagvol}
\ee    
Special Lagrangians are not classified topologically and a given homology class can contain any number of Special Lagrangian manifolds which makes them difficult to study in the case of CYs where we lack an explicit metric. However there is a well known way to construct them as follows. Consider an isometric anti-holomorphic involution $\sigma$ acting on the CY.  This acts as
\be
\sigma\left(J\right) = -J \;,\;\; \sigma\left( \Omega \right) = \overline{\left( e^{i\theta} \Omega \right)} \;. \label{sigjo}
\ee
This follows since the anti-holomorphic property implies $\sigma(I)=-I$, where $I$ is the complex-structure, and the isometric property is $\sigma(g)=g$ with $g$ the metric. The Kahler form is $J=Ig$. There is a possible rotation of $\Omega$ since $g$ only defines $\Omega$ up to a phase which corresponds to Kahler transformations. So the fixed locus of $\sigma$ is a special Lagrangian sub-manifold. Such involutions can be explicitly found given a CY polynomial. Indeed any symmetry of the co-ordinates of the CY combined with complex-conjugation will form an anti-holomorphic (but not necessary isometric) involution as long as 
\be
P_A\left( \sigma(z_i) \right) = \bar{P}_A \left( z_i \right) \;,
\ee
which translates to a condition on the values of the complex-structure moduli being real\footnote{This condition can be thought of as the field truncation from the $\N=2$ hypermultiplets to the $\N=1$ chiral multiplets induced by an orientifold.}. 

A generic anti-holomorphic involution will not be isometric. However we now show that given a set of holomorphic transformations that form a finite representation of some symmetry group (in our case this will be rotations and permutations of the ambient space co-ordinates) it is always possible to go to a point in Kahler moduli space, within the Kahler cone, where for each holomorphic transformation there is a corresponding isometric anti-holomorphic involution. 

We consider a CY manifold $M$ with metric $g$, complex structure $I$ and Kahler form $J$.  The requirement of a positive definite metric is equivalent to having the Kahler form inside the Kahler cone and can be written as
\be
J \left( v, Iv \right) > 0 \;,  \label{kahcon}
\ee
where $v$ is any vector $v \in TM$. Consider the action of an anti-holomorphic involution symmetry $\sigma^*$ of $M$ (for example complex conjugation). The constraint (\ref{kahcon}) implies
\be
\sigma^*\left( J \right) \left( \sigma^*(v) , \sigma^* \left( Iv \right) \right) = - \sigma^*\left(J\right) \left( \sigma^*(v) , I \sigma^*(v)  \right) > 0 \;. \label{sigjkah}
\ee
So $-\sigma^*\left(J\right)$ is also a Kahler form within the Kahler cone and so is
\be
\hat{J} = \half \left( J -  \sigma^*\left(J\right) \right) \;.
\ee
Given a set of holomorphic symmetries of $M$ denoted by a finite representation $g_I$ where $I$ runs over the elements. We have in mind here rotations and permutations of the co-ordinates. Then we have that
\be
\sigma^* g_I = g^{-1}_I \sigma^* \;.
\ee
Now define also anti-holomorphic elements $\tilde{g}_I = \sigma^* g_I$ and consider the form 
\be
\tilde{J} = \sum_I g_I \hat{J} \;, \label{jtil}
\ee 
which also lies inside the Kahler cone by a similar argument to (\ref{sigjkah}). Then
\be
\tilde{g}_J \tilde{J} = \sum_I \sigma^* g_J g_I \hat{J} = \sum_I \sigma^* g_I \hat{J} = \sum_I g_I \sigma^* \hat{J} = -\tilde{J} \;,
\ee 
where the sum over $I$ runs over all the elements. Then choosing the Kahler form to be $\tilde{J}$ implies that all the anti-holomorphic involutions $\tilde{g}_I$ are also isometric. Choosing $\tilde{J}$ can be thought of as taking anti-symmetric combinations of the Kahler form so as to project out the even, under any of the involutions, elements of $h^{(1,1)}$. Physically it is like setting the sizes of some cycles to be equal. We do not address in this paper the question of moduli stabilisation with respect to picking $\tilde{J}$. 

We have now shown that a collection of anti-holomorphic involution symmetries of a CY manifold allows us to identify a set of special Lagrangian submanifolds. In this paper the anti-holomorphic involutions we use are rotations combined with complex conjugation. It is also possible to use permutations of co-ordinates combined with complex conjugation but we leave this for future study \cite{paltifuture}.

\subsection{Anti-holomorphic involutions from rotations}

Consider the following anti-holomorphic involution
\be
\sigma\left( z_i \right) = \overline{\left( \omega_i z_i \right)} \;.\;\;\;\mathrm{(No\; sum)} \label{sigrot}
\ee
Here the rotation angles $\omega_i$ are roots of unity that are symmetries of the CY by themselves
\be
P_A\left( \omega_i z_i \right) = P_A \left( z_i \right) \;.
\ee
So for the quintic they would be fifth roots of unity. The fixed points locus of this involution is 
\ba
P_A\left( z\right) &=& 0 \;, \nn \\
\im{\left( \om_i^{\frac12} z_i \right)} &=& 0 \;.  \label{slageqrot}
\ea
These equations should be thought of as equations on equivalence classes of the homogeneous co-ordinates $\left[ z_i \right]$. So sets of angles related by an overall weighted rotation are equivalent\footnote{We also have the symmetries of the CY $z_i \rightarrow \omega_i z_i$ which relate $\sigma\left(z_i\right)=\bar{\omega}_i \bar{z}_i$ and $\sigma'\left( z_i \right) = \left( \bar{\omega}'_i \right)^2 \bar{\omega}_i \bar{z}_i$. This does not mean that the configurations are equivalent just that they are the same manifold rotated. This rotation is non-trivial once more than one such manifold is considered.}.
These equations define a Special Lagrangian sub-manifold of the CY. Explicitly, on a CY, $\Omega$ takes the form of the residue of the integral of
\be
\frac{\epsilon^{i_{1}i_{2}...i_{n}} Z_{i_{1}} dZ_{i_{2}} \w ... \w dZ_{i_{n}} }{P_{1}P_{2}...P_{K}} \label{genomega}
\ee
around the $K$ circles enclosing the loci $P_A=0$.  Then under $\sigma$ it transforms as (\ref{sigjo}) with 
\be
e^{i\theta_{\sigma}} = \prod_i \om_i \;.
\ee

The set of rotations $\left\{ \omega_i \right\}$ define the special Lagrangian but they do not fix its orientation. This is because the volume form (\ref{slagvol}) involves a square-root which leaves a sign ambiguity. Since the orientation of the cycle is important for our purposes it is more useful to think of a set of rotation angles as leading to two distinct cycles that are the two orientations. We therefore fix the orientation by taking the convention that the square-roots of the rotation angles are taken in the principle branch and we treat the two orientations as two separate cycles. So a special Lagrangian cycle is denoted as
\be
\Pi_{\sigma} = \left\{ \omega_i \right\}_p \;,   \label{slagrot}
\ee
with $p=\pm 1$ denoting the orientation. Note that the transformation $\omega_i \rightarrow \omega_i e^{2\pi i n_i}$ can change the square-root branches and so we must also transform $p \rightarrow p\; \Pi_i (-1)^{n_i} $. \footnote{In \cite{Brunner:1999jq} the issue of the orientation did not arise because the anti-holomorphic involutions studied were of the form $\sigma\left( \omega_i z_i \right) = \overline{\left( \omega_i z_i \right)}$ which are squares of the involutions we consider. In the case of the quintic the squares of the fifth roots of unity and the fifth roots of unity are related by $2\pi$ transformations and so these involutions covered the full spectrum of special Lagrangians. However for the other CY cases studied in this paper this is not the case.}

We can be more specific by restricting to Fermat CY manifolds, which form the bulk of the examples studied in this paper. We can write the defining polynomial as
\be
P\left(z_i\right) = \sum_i \eta^0_i z_i^{h_i} \;,
\ee
where $\eta^0_i = \pm 1$. Here we have a (possibly weighted) projective ambient space $CP^4_{[w_i]}$. We denote the sum of the weights 
\be
d = \sum_i w_i \;.
\ee 
Then $h_i = \frac{d}{w_i}$. We also denote the $d^{\mathrm{th}}$ root of unity 
\be
\alpha = e^{\frac{2\pi i}{d}} \;, 
\ee
so that the rotation angle symmetries are $\omega_i = \alpha^{w_i k_i}$ where $k_i = 0,...,h_i$. We can keep track of the cycle orientation by always writing 
\be
\om_i = \alpha^{m_i} \;,\;\;  0 \leq m_i < d \;. \label{orientmi}
\ee
This means that if we were to send  $\omega_i \rightarrow \omega_i e^{2\pi i n_i}$ then $m_i$ would not be within the range $\{0,d\}$ and so in order to stick to the notation of (\ref{orientmi}) we need to send $m_i \rightarrow m_i - n_i d$ and the orientation $p \rightarrow p\; \Pi_i (-1)^{n_i} $.

The $\eta^0_i$ are coordinate choices but they are useful tools for keeping track of supersymmetry as discussed in section \ref{sec:orient}. The special Lagrangian is given by equations (\ref{slageqrot}) which can be written as a submanifold
\be
\sum_i \eta_i^0 \eta_i \xi_i^{h_i} = 0 \;, \label{slagxieq}
\ee
of an $RP^4_{[w_i]} \subset CP^4_{[w_i]}$ spanned by the real coordinates $\xi_i = \re{\left( \omega_i^{\half} z_i \right)}$. Here $\eta_i = \omega_i^{\frac{h_i}{2}} = \pm 1$. The topology of the manifold depends on the powers $h_i$. If one of them is odd, say $h_1$, then we can map (\ref{slagxieq}) to $RP^3$ by solving for $\xi_1^{h_1}$ and taking the unique real $h_1^{\mathrm{th}}$ root \cite{Brunner:1999jq}. If all the powers are even, which in this paper occurs for the case $CP^4_{[4,1,1,1,1]}$ the topology depends on the $\eta$s and is given in \cite{Roiban:2002iv}.  Note that with the exception of a single case on $CP^4_{[4,1,1,1,1]}$, discussed in \cite{Roiban:2002iv}, where the topology is $\left( S^1 \times S^2 \right)/\mathbb{Z}_2$, the Special Lagrangians are all rigid. 

\subsection{Summing over patches and angles}
\label{sec:sumpat}

The equations defining the special Lagrangian submanifolds (\ref{slageqrot}) are given in terms of homogeneous co-ordinates. In practical calculations it is useful to work patch-wise with affine co-ordinates. This is particularly important when counting intersections between cycles as we discuss in section \ref{sec:intsplag}. Fixing the homogeneity differs for normal projective spaces and weighted projective spaces. We consider the normal case first. 

The way we choose to work on patches is to use the homogeneous rescaling parameter to fix one of the homogeneous co-ordinates, say $z_{j_p}$,  to unity. We label the patch by the choice of the co-ordinate, so that on the patch $P_{j_p}$ we choose $\lambda = z_{j_p}^{-1}$ and work with the affine coordinates
\be
x_i = \frac{z_i}{z_{j_p}} \;,
\ee
so that $x_{j_p}=1$.
This is only possible in a patch where $z_{j_p} \neq 0$ and so we have to sum over all the $z_{j_p}$s in order to cover the whole manifold. It is important in order not to over count intersections that the patches do not overlap and so we use the following scheme. Consider a single homogeneous space $CP^n$. It is spanned by homogeneous co-ordinates
\be
\left\{ z_1, z_2, z_3, ... , z_n  \right\} \;. \label{patchstart}
\ee     
Then we start from the patch $x_{j_p}=x_1=1$
\be
\left\{ 1, x_2, x_3, ... , x_n  \right\} \;. \label{patch1}
\ee    
This covers the manifold apart from a $CP^{n-1}$ spanned by
\be
\left\{ 0, z_2, z_3, ... , z_n  \right\} \;.
\ee     
Now we want to study this patch so we consider the patch
\be
\left\{ x_1, 1 , x_3, ... , x_n  \right\} \;,
\ee 
but restrict $x_1=0$ 
\be
\left\{ 0, 1, x_3,  ... , x_n  \right\} \;. \label{patch2}
\ee   
This will cover the rest of the manifold except a $CP^{n-2}$ spanned by
 \be
 \left\{ 0, 0 , z_3, ... , z_n  \right\} \;.  \label{patchend}
 \ee
 This is then repeated until all the patches are covered. 
 
We note that Identifying the special Lagrangian sub-manifold by the rotation angles as in (\ref{slagrot}) carries a redundancy since two different rotation angle sets that are related by a rotation of all the angles, which is a subset of the complex homogeneous co-ordinate rescaling of the ambient space, should be identified
\be
\Pi = \left\{ \omega_i \right\}_p  \sim \left\{ \alpha^{n} \omega_i \right\}_{p''} \;,\;\; n=0,...,d\;. \label{alor}
\ee
Choosing a patch picks out one representative (which one is our choice) of this equivalence class since we fix $\lambda$. Here the orientation changes according to which branch the angles are in. There are two effects that must be considered here. First the rotations give $p' =p\left( -1 \right)^n$ from changing the orientation $e^{i\theta} \rightarrow e^{i\theta + 2\pi i n}$. Also the  $\omega_i$ are always in the primary branch, as in (\ref{orientmi}), but rotating them with the $\alpha^n$ can take them to a different branch which we then must undo by rotation the angle by $2\pi$ which gives another minus sign for each such angle. 

The case where the ambient space is a weighted projective space is more complicated. Now fixing $\lambda$ carries a remaining symmetry. A useful way to think about this is to think of a weighted projective space with weights $w_i$ as an orbifold of a normal projective space 
\be
CP^n_{[w_i]} = \frac{CP^n}{\Pi_{i} \; \mathbb{Z}_{w_i}} \;. \label{wepro}
\ee
The simple argument to show this can be found in \cite{Hubsch:1992nu} for example. It states that we can map $CP^n(t)$ to $CP^n_{[w_i]}(z)$ by $t_i \ra z_i=t_i^{w_i}$. This is bijective if identify $t_i \sim \alpha_{i}t_i$ where $\alpha_{i}$ is the $w_i$th root of unity. This orbifolding manifests itself locally patchwise; in order to work on a patch $P_{j_p}$ we should take 
\be
\lambda^{w_{j_p}} = z_{j_p}^{-1} \;\; \ra \;\; \lambda =\left( \alpha_{j_p} \right)^{k_{j_p}} \left(z_{j_p}\right)^{-\frac{1}{w_{j_p}}} \;, 
\ee
where $\alpha_{j_p}$ is a $w_{j_p}$th root of unity and $k_{j_p}$ is a free integer ranging up to $w_{j_p}$. So now fixing the patch still leaves a ${\mathbb Z}_{w_{j_p}}$ rotation freedom in the affine coordinates
\be
x_i = z_i \left(\alpha_{j_p}\right)^{k_{j_p} w_i} \left(z_{{j_p}}\right)^{-\frac{w_i}{w_{j_p}}} \;. \label{affwe}
\ee
This can be thought of as a local orbifold. 
To take this into account, when we work on a patch we need to sum over the $k_{j_p}$s so that on a given patch $P_{j_p}$ the cycle is given by the sum over sets of rotation angles
\be
\Pi^{P_{j_p}} = \sum_{k_{j_p}} \left\{ \left(\alpha_{j_p}\right)^{k_{j_p} w_i} \om_i \right\}_{p(k_{j_p})} \;. 
\ee
To sum over the patches we can still use the technique outlined in (\ref{patchstart})-(\ref{patchend}) as long as the CY manifold is smooth which means that the singularities of the ambient space do not affect the intersections. 

It is important to note that the constraint $\im{\xi_i} = 0$ on the homogeneous co-ordinates does not imply the same for the affine coordinates. From (\ref{affwe}) we see that $\im{x_i}$ need not vanish. On patch $P_{j_p}$ the $x_i$ take values in $\mathbb{R} \times \left\{ \exp\left(\frac{n \pi i w_i }{w_{j_p}}\right) \right\}$ for $n=1,..,w_{j_p}$. Of this set we should identify values related by $\left(\alpha_{j_p}\right)^{k_{j_p} w_i} = \exp\left(\frac{2 k_{j_p} \pi i w_i }{w_{j_p}}\right)$ and so we are left with 
\be
x_i \in \mathbb{R} \times \left\{ 1, e^{\frac{w_i \pi i}{w_{j_p}}} \right\} \;.
\ee
This is important in counting solutions as in section \ref{sec:intsplag}. 

\subsection{Counting special Lagrangian submanifolds}
\label{sec:countslags}

We are interested in finding substantial sets of special Lagrangian submanifolds of explicit CY examples. In this paper we restrict ourselves to smooth CYs (which do not require singularity blow-ups), which include all the complete intersection CYs in (products of) normal projective spaces and the weighted projective spaces cases in table \ref{wcyexa}. We also restrict to special Lagrangians that are fixed points of rotations only and leave permutations for future work \cite{paltifuture}. 
The method of identifying special Lagrangian submanifolds discussed in the previous sections requires identifying rotation symmetries of the CY. Finding a large set of these is easiest in the case of weighted projective spaces and so we will primarily restrict ourselves to those cases. The symmetries can easily be read off the explicit polynomial forms chosen in table \ref{wcyexa}. In the table we display the number of distinct special Lagrangian submanifolds that can be constructed in each case. For example, consider the quintic. We have a $Z_5^5$ symmetry group acting as rotations of the co-ordinates by fifth roots of unity. A $Z_5$ subgroup of that is trivial, in that it is part of the complex homogeneous rescaling symmetry, leaving $625$ distinct cycles. We denote the set of special Lagrangians that have a vanishing calibration angle supersymmetric to mark the fact that they all preserve the same supersymmetry. In constructing models we only use members of this set.\footnote{Of course we could have chosen a different, non-vanishing, angle to pick out a different supersymmetric set. This would correspond to different choices for the $\eta^0_i$s as explained in section \ref{sec:orient}.} In table \ref{wcyexa} we also include the Hodge numbers of the manifolds, which were calculated in \cite{Klemm:1992tx}, and the rank of the intersection matrix of the special Lagrangians which can be calculated using the techniques of section \ref{sec:intsplag}. Note that the rank of the intersection matrix is smaller than $b^3$ for all the cases apart from the quintic. This means that the set of special Lagrangian manifolds do not span the full homology of the manifold.  
\begin{table}
\center
\begin{tabular}{|c|c|c|c|c|c|c|}
\hline
Ambient Space & Defining Polynomial & SLAG & SUSY & $b^3$ & Rank \\
\hline
$CP^4_{[1,1,1,1,1]}$ & $ P = \eta^0_1 z_1^5 + \eta^0_2 z_2^5 + \eta^0_3 z_3^5 + \eta^0_4 z_4^5 + \eta^0_5 z_5^5 = 0$ & $625$ &  $125$ & $204$ & $204$\\  
\hline
$CP^4_{\left[ 2,1,1,1,1 \right]}$ &$ P =\eta^0_1  z_1^3 + \eta^0_2  z_2^6 + \eta^0_3 z_3^6 + \eta^0_4  z_4^6 + \eta^0_5  z_5^6 = 0$ & $648$ &  $108$ & $208$ & $54$\\ 
\hline
$CP^4_{\left[ 4,1,1,1,1 \right]}$ &$ P = \eta^0_1 z_1^2 +\eta^0_2  z_2^8 + \eta^0_3  z_3^8 + \eta^0_4  z_4^8 + \eta^0_5  z_5^8 = 0$ & $960$ &  $120$ & $300$ & $64$\\
\hline
$CP^4_{\left[ 5,2,1,1,1 \right]}$ &$P = \eta^0_1 z_1^2 + \eta^0_2 z_2^5 + \eta^0_3  z_3^{10} + \eta^0_4  z_4^{10} + \eta^0_5  z_5^{10} = 0$ & $1000$ & $100$ & $292$ & $100$\\
\hline
\end{tabular}
\caption{Smooth CYs within weighted projective spaces and the quintic.}
\label{wcyexa}
\end{table}

\section{Intersecting special Lagrangian submanifolds}
\label{sec:intsplag}

In model building, the chiral spectrum is determined by the intersection numbers of the cycles wrapped by the branes \cite{Berkooz:1996km}. In this section we study counting intersections between pairs of special Lagrangian submanifolds that are constructed using the methods of section \ref{sec:fslags}. We denote an intersection supersymmetric if the two special Lagrangian submanifolds are calibrated with the same angle. 
Given two distinct special Lagrangian submanifolds they can intersect on loci of dimensions zero, one or two. In section \ref{sec:ponint} we consider point intersection an discuss higher dimensional intersection in section \ref{sec:intcur} and appendix \ref{sec:curandsur}. In appendix \ref{sec:intexam} we present an explicit computation of intersections between two special Lagrangian manifolds which serves as a clarifying example. 

\subsection{Point intersections}
\label{sec:ponint}

The counting of point intersections of two special Lagrangians corresponds to simply counting common solutions to their defining equations. There is also a sign associated to the orientation of each intersection which we return to soon.  Consider two special Lagrangians denoted by
\be
\Pi_1 = \left\{ \om'_i \right\}_{p'} \;,\;\; \Pi_2 = \left\{ \tilde{\om}_i\right\}_{\tilde{p}} \;.
\ee 
We can always redefine our co-ordinates $z'_i = \om'_i z_i$ and pick the relative orientation so that solving for the intersections is the same as solving the system
\be
\Pi_1 = \left\{ 1 \right\}_{+}  \;,\;\; \Pi_2 = \left\{ \om_i \right\}_{p} \;, \label{intor}
\ee 
where $\om_i  = \tilde{\om}_i \bar{\om}'_i$. Note that here we are calculating $\Pi_1 \cdot \Pi_2$ and so we perform the co-ordinate change so that $\Pi_1 \rightarrow \left\{ 1 \right\}_{+}$. If we were calculating $\Pi_2 \cdot \Pi_1$ we would perform a co-ordinate change so that $\Pi_2 \rightarrow \left\{ 1 \right\}_{+}$ which would give the opposite intersection number.
Recall that we keep track of the cycle orientation by always writing the rotation angles in the primary patch as in (\ref{orientmi}). If we write 
\be
\om'_i = \alpha^{m'_i} \;\;\,\; \tilde{\om}_i = \alpha^{\tilde{m}_i} \;\;\,
\ee 
then the orientation $p$ is calculated as 
\be
p = p' \tilde{p} \;\prod_i (-1)^{p_i} \;,
\ee
where $p_i=0$ if $m'_i\leq\tilde{m}_i$ and  $p_i=1$ if $m'_i>\tilde{m}_i$. 
The point intersections are given by the number of point solutions to the set of equations
\ba
P_A(\re{(z_i)}) &=& 0 \;, \label{int1} \\
P_A(\re{(\om^{\frac12}_i z_i)}) &=& 0 \;, \label{int2}\\
\im{z}_i &=& 0 \;,\label{int3} \\
\im{(\om^{\frac12}_iz_i)} &=& 0 \;.\label{int4}
\ea
Equations (\ref{int3}) and (\ref{int4}) imply that if $\om_i \neq 1$ then $z_i=0$. Else the equations are equivalent and the real part of $z_i$ is unconstrained. From here on when we refer to a rotation we consider only non trivial ones $\om_i \neq 1$. 
Imposing (\ref{int3}) and (\ref{int4}) implies that (\ref{int1}) and (\ref{int2}) are equivalent. Then point intersections can only occur when $\Pi_2$ has three non-trivial rotation angles, which will set the three rotated co-ordinates to zero. The intersection number is then given by the number of distinct solutions to the remaining equation (\ref{int1}). This is a local intersection number  $I^{(L)}_{\Pi_1\Pi_2}$   since it depends on the $\omega_i$ which depend on the patch. 

The sign of the local intersection can be computed as the sign between the orientation of the CY manifold and the orientation induced by the tangent bundles of the two special Lagrangian submanifolds. This sign is given by
\be
\mathrm{sgn} \; I^{(L)}_{\Pi_1\Pi_2}\left( \om_i \right) = \mathrm{sgn} \; \left[p_1 p_2 \;\frac{\re{\left( e^{i\frac{\theta_{\Pi_1}}{2}}\Omega \right)}\wedge\re{\left( e^{i\frac{\theta_{\Pi_2}}{2}} \Omega \right)}}{i\Omega \wedge \bar{\Omega} } \right]\;.
\ee
The division just means take out the volume form from both numerator and denominator. There is an overall sign ambiguity which would flip the sign of all the intersections that is just equivalent to what we call left handed or right handed. The important thing is the dependence of the data of the individual cycles which is their specified rotation angles and orientations $p_1$ and $p_2$. For the intersection (\ref{intor}) this reads
\be
\mathrm{sgn} \; I^{(L)}_{\Pi_1\Pi_2}\left( \om_i \right) = \mathrm{sgn} \; \left[  p \; \prod_i \; \im{\left(\om^{\half}_i\right)} \right] = p \;,
\ee
since the rotation angles are all in the primary branch. 

The global total intersection number corresponds to summing over the rotation angles and patches as described in section \ref{sec:sumpat} so as to cover the full cycle. So summing over the rotation angle sets gives the intersection number for that patch
\be
I^{(P_{j_p})}_{\Pi_1\Pi_2} = \sum_{k_{j_p}}  I^{(L)}_{\Pi_1\Pi_2}\left( \alpha_{j_p}^{k_{j_p} w_i} \om_i \right) \;, \label{suminpatch}
\ee
where $P_{j_p}$ denotes the patch where $x_{j_p}=1$ and we recall that $\alpha_{j_p}$ is the $w_{j_p}$th root of unity and $k_{j_p}$ is a free integer ranging up to $w_{j_p}$. The total intersection number is given by summing over the patches 
\be
I_{\Pi_1\Pi_2} = \sum_i  I^{(P_i)}_{\Pi_1\Pi_2} \;. 
\ee

\subsection{Intersections on curves and surfaces}
\label{sec:intcur}

Two special Lagrangian submanifolds can also intersect on loci of dimensions one (curves) and two (surfaces). There is still a relevant intersection number associated with these cases that is the self-intersection of the intersection locus. To calculate this we need to know the topology of the intersection locus which in turn depends on the explicit form of the polynomials defining the special Lagrangians. Therefore this essentially needs to be done on a case-by-case basis. In appendix \ref{sec:curandsur} we perform this analysis for the three weighted projective spaces in table \ref{wcyexa}.

There is an important general property of surface intersections: two special Lagrangians that are calibrated by the same phase never intersect on a surface. This follows simply from the fact that such an intersection requires the two special Lagrangians to have only one rotation angle different which means they can not be calibrated by the same angle. This implies that in supersymmetric model building, where all the branes are wrapping cycles with the same calibration phase, surface intersections do not play a role in the matter spectrum calculation. For this reason their discussion has been relegated to the appendix. Surface intersections are nonetheless important for calculating the rank of the intersection matrix of the full special Lagrangian set as displayed in table \ref{wcyexa}. This, in turn, is important to know for addressing homological issues such as tadpole cancellation for which the special Lagrangian set needs to span the full homology of the manifold. Therefore they still play a role (though in the examples studied a minor one since the special Lagrangians do not span the full homology).

The only closed one-dimensional manifold is the circle which has vanishing self intersection. Therefore curve intersections are always vanishing. 

\section{Intersecting Branes}
\label{sec:intbran}

So far our discussion has concentrated on the geometry of special Lagrangian submanifolds. In this section we discuss the physics associated to wrapping D6-branes and O6-planes on these submanifolds. This is largely a review and the results used are well documented in the literature \cite{Blumenhagen:2006ci} and so we will be brief and simply state them. The main aim of this section is to formulate the conditions in a form that is suitable for use in section \ref{sec:modbuild} where we study explicit models. 

Throughout this section we use three types of branes that can appear in a given model. The set $\{\mathrm{visible}\}$ corresponds to branes whose gauge group is part of the gauge group that forms the visible sector gauge group. The set $\{\mathrm{exotic}\}$ corresponds to branes whose gauge group is not part of the visible gauge group but are needed in order to cancel tadpoles. In a given model we specify these two sectors explicitly. Finally the set $\{\mathrm{hidden}\}$  corresponds to branes that are needed for consistency conditions such as tadpoles or for a massless $U(1)$ for which we can not identify the appropriate special Lagrangian to wrap. In analogy to local models these type of branes can be thought of as bulk branes. 

\subsection{Supersymmetry}

Branes and orientifold planes wrapping Special Lagrangian submanifolds preserve half the supersymmetry of the background CY and the angle $\theta_{\Pi}$ gives the linear combination of the supersymmetry spinors that is preserved \cite{Becker:1995kb}. In order to preserve this remaining ${\cal N}=1$ supersymmetry completely all branes and orientifold planes must be calibrated with the same angle, if any branes are calibrated with a different angle supersymmetry is completely broken. However the phenomenology of supersymmetry breaking depends strongly on which branes are non-supersymmetric. If the branes that give rise to the visible sector break supersymmetry then the visible scale of supersymmetry breaking is the string scale and this is only compatible with a solution to the hierarchy problem if the string scale is near the TeV scale. Unfortunately in the case of the quintic this is the only possibility as special Lagrangians that are calibrated with the same angle do not intersect \cite{Blumenhagen:2002wn}. Perhaps a more attractive possibility, and one that we restrict to in this paper, is having the supersymmetry breaking sector decoupled from the visible sector. This means that branes that break supersymmetry do not intersect the visible sector (at least no net intersection number). Supersymmetry breaking is then mediated gravitationally and potentially also through gauge mediation. In this case the scale of visible supersymmetry breaking is not tied to the string scale but rather fixed dynamically and is a question of moduli stabilisation. So to summarise, the models we consider are ones where supersymmetry is preserved by the visible sector and is possibly broken by a hidden sector that does not intersect the visible one.  

\subsection{Orientifolds}
\label{sec:orient}

All the models we study contain orientifold planes which are important for model building and tadpole cancellation purposes. These are O6 planes wrapping special Lagrangian submanifolds that are fixed loci of a singled-out anti-holomorphic isometric involution. The orientifolds wrap all the fixed loci of this involution. In all our examples we choose this involution to be simply complex-conjugation.  Then the orientifold wraps the cycle given by all the rotation angles being unity 
\be
\Pi_O = \left\{ 1 \right\}_+ \;.
\ee
Actually it is possible that there are other cycles that are fixed loci of this involution but just are not generated by the methods we are considering. Indeed we generically expect many such fixed loci. Some other fixed loci can be generated. For example on the quintic the fixed loci also include fixed points of permutations of two pairs of coordinates. This gives 16 identifiable cycles wrapped by O6 planes. In the mirror IIB picture these would be a single O3 plane and 15 O7 planes.

It is possible to associate a different anti-holomorphic involution with the orientifold by also including a rotation. However this can always be turned to just complex conjugation by an appropriate coordinate change. The result of this coordinate change is to simply change the $\eta^0_i$ in table \ref{wcyexa}. Therefore by allowing the $\eta^0_i$ to vary we are allowing for the different orientifold cycles. Each choice of $\eta^0_i$ corresponds to preserving a different $\N=1$ supersymmetry. 

Apart from modifying the tadpole equations the orientifolds imply that every brane has an orientifold image. We denote the orientifold brane image by a prime and, in our case where the orientifold involution is complex conjugation, it is simply obtained by complex conjugating the rotation angles that defined the original cycle
\be
\Pi_a = \left\{ \om_i \right\}_+ \implies \Pi_{a'} = \left\{ \bar{\om}_i \right\}_+ \;.
\ee

\subsection{Chiral Spectrum}

The massless chiral spectrum of particles that are end-points of strings stretching between two branes is given by the topological intersection number of the special Lagrangian submanifolds wrapped by the branes. We denote the intersection numbers as
\ba
\Pi_a \cdot \Pi_b &\equiv& I_{ab} = -I_{ba} = I_{b'a'} = -I_{a'b'} \;, \\
\Pi_a \cdot \Pi_{b'} &\equiv& I_{ab'} = I_{ba'} = -I_{a'b} = -I_{b'a} \;.
\ea 
The spectrum of fields and representations for strings stretching between stack $a$ of $N_a$ branes and stack $b$ of $N_b$ branes is given in table \ref{inttab} \cite{Berkooz:1996km, Blumenhagen:2002wn}.
\begin{table}
\center
\label{inttab}
\begin{tabular}{|c|c|}
\hline
Multiplicity ($\times$ chirality) & Representation \\
\hline 
\raisebox{-.3ex}{$I_{ab}$} & \raisebox{-.3ex}{$\left(\bar{N}_a,N_b\right)$} \\
\hline
\raisebox{-.3ex}{$I_{a'b}$} & \raisebox{-.3ex}{$\left(N_a,N_b\right)$} \\
\hline
\raisebox{-.3ex}{$\half \left( I_{a'a} + I_{O6,a} \right)$} & $A_{N_a}$ \\
\hline
\raisebox{-.3ex}{$\half \left( I_{a'a} - I_{O6,a} \right)$} & $S_{N_a}$ \\
\hline
\end{tabular}
\caption{Chiral particle spectrum for brane stacks $a$ and $b$ with $N_a$ and $N_b$ branes respectively. $\left(N_a, N_b\right)$ denotes the bi-fundamental representation and $S$ and $A$ denote the symmetric and anti-symmetric representations respectively. }
\end{table}
Here we use the conventions that the particles are left handed Weyl fermions. The intersection numbers are topological and count the net chirality. There are also vector pairs that arise at local pairs of oppositely oriented intersections. We usually suppress these in listing the chiral content of a model, it is expected that these fields become massive at the string scale since they are not protected by chirality. 

\subsection{Tadpoles and anomalies}

There are strong consistency constraints coming from cancellation of Ramond-Ramond (RR) tadpoles. The condition for this is \cite{Berkooz:1996km,Blumenhagen:2002wn,Blumenhagen:2006ci}
\be
\sum_{a} N_a \left( \Pi_a + \Pi_{a'}  \right) - 4 \Pi_{O6} = 0 \;. \label{rrtad}
\ee
We can also consider the weaker constraint, which is closely related to anomaly cancellation, 
\be
\sum_{a} N_a \left( I_{ab} + I_{a'b} \right) - 4 I_{O6,b} = 0 \;\;\; \forall \; b \;\in \{\mathrm{visible}\} \;. \label{rrano}
\ee
Recall the set $\{\mathrm{visible}\}$ denotes the cycles wrapped by the branes that contribute to the visible gauge group.
This is weaker than the tadpole constraint since the cycles wrapped by branes need not form a complete homological basis. However as long as (\ref{rrano}) is satisfied we can always add a hidden sector to satisfy (\ref{rrtad}) and this hidden sector will be truly hidden in that there will be no (chiral) massless states charged under the hidden and visible sector. The hidden sector couples gravitationally and through massive states to the visible sector and can serve as a source of supersymmetry breaking.\footnote{There are vector-like states that couple the hidden sector to the visible one. In stating that mediation is through massive states we assume they gain a large mass.} It is important to note that the hidden sector need not break supersymmetry at all.

The scenario described in the previous paragraph is particularly relevant for our examples. This is because in our examples the special Lagrangian submanifolds do not span the full homology of the manifold. The requirement to span the full homology is that the rank of the intersection matrix of the special Lagrangian submanifolds is equal to $b^3$.  
As can be seen in table \ref{wcyexa} this only occurs for the quintic. Since we are unable to span the full homology it is not possible to determine, solely using intersection numbers, if a sum of special Lagrangians is homologically trivial or not. The best we can do is make sure we satisfy the weaker constraint (\ref{rrano}) which guarantees an anomaly free theory with a truly hidden sector. 

This situation is analogous to local models of intersecting branes where a lack of knowledge of the bulk completion of the model implies that global tadpoles must be canceled by a hidden sector. In this way our models are similar in spirit to local models. 

\subsection{Massless $U(1)$s}

Any $U(1)$ gauge fields in the spectrum can become massive through the Green-Schwartz mechanism \cite{Berkooz:1996km}. If the $U(1)$ is anomalous (with respect to the open string spectrum), and the RR tadpoles are satisfied, this is guaranteed. If it is not anomalous it will stay massless if it is homologically trivial and so not coupled to the closed string RR fields. By this we mean that a $U(1)$ defined as the linear combination
\be
U(1) = \sum_{a} Q_a U(1)_a \;,
\ee 
will remain massless if  \cite{Blumenhagen:2002wn}
\be
\sum_a N_a Q_a \left( \Pi_a - \Pi_{a'} \right) = 0 \;.   \label{masslessu1}
\ee
Here we face the same problem as with the tadpoles; the constraint (\ref{masslessu1}) is homological and we can not guarantee to satisfy it if we do not have a complete homological basis of cycles to work with. 

We can bypass this problem, in a similar way to that used for the tadpoles, by introducing a hidden sector to make a chosen linear combination of the $U(1)$s massless. As long as this hidden sector does not intersect any other branes there will be no new chiral matter charged under the visible sector gauge fields. So the `local'-type condition on massless hypercharge is 
\be
\sum_a N_a Q_a \left( I_{ab} - I_{a'b} \right) = 0 \;,\;\; \forall \;b \;\in \{\mathrm{visible+exotic}\} \;, \label{hyploc}
\ee
where the set $\{\mathrm{visible+exotic}\}$ needs to include all the branes in the construction.

It is possible to check that a $U(1)$ is massive by finding a cycle $b$ such that
\be
\sum_a N_a Q_a \left( I_{ab} - I_{a'b} \right) \neq 0 \;. \label{massiveu1}
\ee
For a $U(1)$ to remain massless (\ref{massiveu1}) must vanish for all $b$.
Indeed by doing this for all the available cycles $b$ we can generate useful constraints on the coefficients $Q_a$. The number of massless $U(1)$s can not exceed the number of brane stacks minus the number of linearly independent constraints on the $Q_a$ generated this way. 

\section{Model building on $CP_{[5,2,1,1,1]}$}
\label{sec:modbuild}

In this section we present two example models, one is a Pati-Salam model and the other MSSM-like. They are both configurations within the CY in the weighted projective space $CP_{[5,2,1,1,1]}$ with the coordinate choice $\eta^0_i=+1$. Recall this CY has 10 sets of 100 supersymmetric cycles and the coordinate choice picks out one set corresponding to that which includes the orientifold cycle $\Pi_0=\left\{0,0,0,0,0\right\}_+$. We label the cycles using the notation of appendix \ref{sec:intexam} in which a cycle is represented by a set of rotation angles (corresponding to an anti-holomorphic involution) which in turn are denoted by integers that are the corresponding power of the 10th root of unity. 

The models require a hidden sector to solve the tadpole constraints and the MSSM-like model of section \ref{sec:mssmlike} requires a hidden sector to guarantee a massless hypercharge. In that sense they can be thought of as analogous to local models. We do not present any analysis of further important details such as Yukawa couplings and only present the massless gauge group and chiral spectrum.

The models were all found using a computer search. The models presented are certainly not unique and there are many others like them, but a quantitative analysis is beyond the scope of this paper. 

We looked for possible GUT models within the weighted projective spaces using a computer search but could not find any models models with three generations of the anti-symmetric representation of $SU(5)$.  

We also studied some CICY spaces, specifically $P[4\|5]$, $P[5\|2\;4]$, $P[5\|3\;3]$, $P[6\|2\;2\;3]$, $P[7\|2\;2\;2\;2]$ and $P\left[
\begin{tabular}{ccccc}
 $3\; \|$ & $3$ & $1$ \\ 
 $2\; \|$ & $0$ & $3$   
\end{tabular}
\right]
$. None of them provided a rich enough intersection matrix to produce any reasonable spectrum.\footnote{There are a further $102$ CICYs that have powers in their polynomial greater or equal to $3$. A systematic study of their intersection matrix is beyond the scope of this paper.}

In appendix \ref{sec:someothermodels} we also construct a supersymmetric Pati-Salam-like model on $CP^4_{[2,1,1,1,1]}$ and a supersymmetric two-generation $SU(5)$ GUT model on $CP^4_{[4,1,1,1,1]}$ as simple examples of models on those manifolds.

\subsection{Supersymmetric Pati-Salam model}
\label{sec:patisal}

The model is constructed from three stacks of branes $\{a,b,c\}$ with $N_a=4$, $N_b=2$ and $N_c=2$. This gives rise to the Pati-Salam gauge group $SU(4)\times SU(2)_L \times SU(2)_R$. The three extra $U(1)$s will be shown to all gain Green-Schwarz masses. The cycles that the branes wrap are
\ba
\Pi_a &=& \left\{0,0,0,3,7\right\}_{-} \;\;,\;\; \Pi_{a'} = \left\{0,0,0,7,3\right\}_{-} \;, \nn \\
\Pi_b &=& \left\{0,0,7,2,1\right\}_{-} \;\;,\;\; \Pi_{b'} = \left\{0,0,3,8,9\right\}_{+} \;, \nn \\
\Pi_c &=& \left\{0,0,3,9,8\right\}_{+} \;\;,\;\; \Pi_{c'} = \left\{0,0,7,1,2\right\}_{-} \;.
\ea
The intersection matrix for these cycles is given in table \ref{psint}.
\begin{table}
\center
\begin{tabular}{|c|ccccccc|}
\hline
\;& $\Pi_a$ & $\Pi_b$ & $\Pi_c$ & $\Pi_{a'}$ & $\Pi_{b'}$ & $\Pi_{c'}$ & $\Pi_0$ \\
\hline
$\Pi_a$ & $0$ & $-1$ & $1$ & $0$ & $-2$ & $2$ & $0$ \\  
$\Pi_b$ & \;  & $0$ & $1$ & \;  & $3$ & $0$ & $1$\\  
$\Pi_c$ & \;  & \;  & $0$ & \;  & \;  & $-3$ & $-1$ \\
\hline
\end{tabular}
\caption{Intersection numbers for Pati-Salam model.}
\label{psint}
\end{table}
The intersection numbers give rise to the field spectrum given in table \ref{pspec}. We only display the chiral spectrum apart from the required vector pair of Higgses needed to break to the MSSM\footnote{The existence of the vector pair is calculated explicitly in appendix \ref{sec:intexam} as the example intersection. The model also contains six other vector-like pairs in different representations.}. We also suppress hidden gauge group charges and the appropriate charges can be read from the intersection matrix \ref{psint}.
\begin{table}
\center
\begin{tabular}{|c|c|c|}
\hline
Field & Multiplicity & Representation \\
\hline
$\mathrm{Q}_{\mathrm{L}}$ & $3$ & $\left( 4,2,1 \right)$ \\  
$\mathrm{Q}_{\mathrm{R}}$ & $3$ & $\left( \bar{4},1,2 \right)$ \\  
h & $1$ & $\left( 1,2,2 \right)$ \\  
$\mathrm{H}_{+}$ & $1$ & $\left( \bar{4},1,2 \right)$ \\  
$\mathrm{H}_{-}$ & $1$ & $\left( 4,1,2 \right)$ \\  
\hline 
\hline 
$\mathrm{B}_{1}$ & $1$ & $[\mathrm{S}]_{SU(2)}$ \\  
$\mathrm{B}_{2}$ & $2$ & $[\mathrm{A}]_{SU(2)}$ \\  
$\mathrm{C}_{1}$ & $1$ & $[\mathrm{S}]_{SU(2)}$ \\  
$\mathrm{C}_{2}$ & $2$ & $[\mathrm{A}]_{SU(2)}$ \\  
\hline
\end{tabular}
\caption{Chiral spectrum of Pati-Salam model. The fields are left handed Weyl fermions.}
\label{pspec}
\end{table}
The spectrum is clean with only two charged chiral exotics denoted $B_1$ and $C_1$.\footnote{Recall that the anti-symmetric representation of $SU(2)$ is a singlet.} It is simple to check that the `local' tadpole conditions are satisfied and so a hidden sector added to cancel global tadpoles does not give rise to any new chiral matter charged under the visible gauge groups. 

To show that all the $U(1)$s are massive we need to prove that there are no non-vanishing $Q_{\{a,b,c\}}$ such that 
\be
\Pi_{U(1)} = N_a Q_a \left( \Pi_a - \Pi_{a'} \right) + N_b Q_b \left( \Pi_b - \Pi_{b'} \right) + N_c Q_c \left( \Pi_c - \Pi_{c'} \right) \;, \label{piu1}
\ee
is homologically trivial. This can be shown by calculating intersections of $\Pi_{U(1)}$ with other cycles which must vanish if it is to be homologically trivial. Intersecting with the three cycles $\{0,0,0,0,0\}_{+}$, $\{0,0,0,0,1\}_{+}$, $\{0,0,0,0,2\}_{+}$ gives the constraints $Q_b=Q_c$, $Q_b=0$, $Q_a=-Q_c$ respectively which imply $Q_a=Q_b=Q_c=0$ and so there are no massless $U(1)$s.

\subsection{MSSM-like model}
\label{sec:mssmlike}

In this section we present an MSSM-like model. This is a 7 brane model. 5 branes are used to construct the MSSM with a `locally' massless hypercharge as in (\ref{hyploc}), and 2 branes are added to satisfy the `local' tadpoles (\ref{rrano}) where the index $b$ runs over the 5 visible sector branes. The point of adding the 2 `exotic' branes is that we can determine the exotic spectrum since any branes added to satisfy global tadpole cancellation will not give rise to matter charged under a visible gauge group. The number of branes wrapping the seven cycles are $N_a=3$, $N_b=2$, $N_c=1$, $N_d=1$, $N_e=1$, $N_f=1$, and $N_g=1$ such that the low energy gauge group is $SU(3)\times SU(2) \times U(1)_H$ where the hypercharge $U(1)_H$ is massless in a `local' sense as discussed below. 

The cycles wrapped by the branes are 
\ba
\Pi_a &=& \left\{0,0,0,3,7\right\}_{-} \;\;,\;\; \Pi_{a'} = \left\{0,0,0,7,3\right\}_{-} \;, \nn \\
\Pi_b &=& \left\{0,0,3,8,9\right\}_{+} \;\;,\;\; \Pi_{b'} = \left\{0,0,7,2,1\right\}_{-} \;, \nn \\
\Pi_c &=& \left\{0,0,3,0,7\right\}_{-} \;\;,\;\; \Pi_{c'} = \left\{0,0,7,0,3\right\}_{-} \;, \nn \\
\Pi_d &=& \left\{0,0,4,1,5\right\}_{-} \;\;,\;\; \Pi_{d'} = \left\{0,0,6,9,5\right\}_{+} \;, \nn \\
\Pi_e &=& \left\{0,0,7,8,5\right\}_{-} \;\;,\;\; \Pi_{e'} = \left\{0,0,3,2,5\right\}_{+} \;, \nn \\
\Pi_f &=& \left\{0,0,2,6,2\right\}_{-} \;\;,\;\; \Pi_{f'} = \left\{0,0,8,4,8\right\}_{+} \;, \nn \\
\Pi_g &=& \left\{0,0,3,4,3\right\}_{-} \;\;,\;\; \Pi_{g'} = \left\{0,0,7,6,7\right\}_{+} \;.
\ea
The intersection matrix for this set of cycles is given in table \ref{mssmint}. The intersection numbers give rise to the chiral spectrum given in table \ref{mssmpec}.
\begin{table}
\center
\begin{tabular}{|c|ccccccccccccccc|}
\hline
\;& $\Pi_a$ & $\Pi_b$ & $\Pi_c$ & $\Pi_d$ & $\Pi_e$ & $\Pi_f$ & $\Pi_g$ & $\Pi_{a'}$ & $\Pi_{b'}$ & $\Pi_{c'}$ & $\Pi_{d'}$ & $\Pi_{e'}$ & $\Pi_{f'}$ & $\Pi_{g'}$ & $\Pi_0$ \\
\hline
$\Pi_a$ & $0$ & $-2$ & $0$ & $1$ & $2$ & $0$ & $1$ & $0$ & $-1$ & $1$ & $3$ & $-1$ & $-1$ & $0$ & $0$\\  
$\Pi_b$ & \;  & $0$  & $-2$ & $1$ & $0$ & $1$ & $-2$ & \;& $-3$ & $-1$ & $1$ & $2$ & $0$ & $-1$ & $-1$\\  
$\Pi_c$ & \;  & \;   & $0$ & $-1$ & $-1$ & $0$ & $2$ & \;& \;  & $0$ & $-1$ & $-2$ & $0$ & $-2$ & $0$\\  
$\Pi_d$ & \;  & \;   & \;  & $0$ & $0$ & $0$ & $-1$  & \;& \;  & \;  & $-2$ & $0$ & $-1$ & $-2$ & $0$\\  
$\Pi_e$ & \;  & \;   & \;  & \;  & $0$  & $0$ & $3$  & \;& \;  & \;  & \;  & $2$ & $-1$ & $2$ & $0$\\  
$\Pi_f$ & \;  & \;   & \;  & \;  & \;   & $0$ & $-1$ & \;& \;  & \;  & \;  & \;  & $1$ & $0$ & $1$\\  
$\Pi_g$ & \;  & \;   & \;  & \;  & \;   & \;  & $0$  & \;& \;  & \;  & \;  & \;  & \;  & $-1$ & $1$\\  
\hline
\end{tabular}
\caption{Intersection numbers for MSSM-like model.}
\label{mssmint}
\end{table}
\begin{table}
\center
\begin{tabular}{|c|c|c|}
\hline
Field & Multiplicity & Representation \\
\hline
Q & $3$ & $\left( 3,2 \right)_{\frac16}$ \\  
U & $3$ & $\left( \bar{3},1 \right)_{-\frac23}$ \\  
D & $3$ & $\left( \bar{3},1 \right)_{\frac13}$ \\  
L & $3$ & $\left( 1,2 \right)_{-\frac12}$ \\  
E & $3$ & $\left( 1,1 \right)_{1}$ \\  
N & $3$ & $\left( 1,1 \right)_{0}$ \\  
$\mathrm{H}_{\mathrm{u}}$ & $1$ & $\left( 1,2 \right)_{\frac12}$ \\  
$\mathrm{H}_{\mathrm{d}}$ & $1$ & $\left( 1,2 \right)_{-\frac12}$ \\  
\hline 
\hline 
$\mathrm{H}_1$ & $1$ & $\left( 1,2 \right)_{\frac12}$ \\  
$\mathrm{H}_2$ & $1$ & $\left( 1,2 \right)_{-\frac12}$ \\ 
$\mathrm{B}_1$ & $1$ & $\left( \bar{3},1 \right)_{-\frac23}$ \\  
$\mathrm{B}_2$ & $1$ & $\left( \bar{3},1 \right)_{\frac23}$ \\  
$\mathrm{B}_3$ & $1$ & $\left( \bar{3},1 \right)_{-\frac16}$ \\  
$\mathrm{B}_4$ & $1$ & $\left( 3,1 \right)_{\frac16}$ \\  
$\mathrm{C}_1$ & $4$ & $\left( 1,2 \right)_{0}$ \\  
$\mathrm{D}_1$ & $7$ & $\left( 1,1 \right)_{\frac12}$ \\  
$\mathrm{D}_2$ & $7$ & $\left( 1,1 \right)_{-\frac12}$ \\  
$\mathrm{E}_{1}$ & $1$ & $[\mathrm{S}]_{SU(2)}$ \\  
\hline 
\hline 
$\mathrm{F}_1$ & $3$ & $\left( 1,1 \right)_{X}$ \\  
$\mathrm{F}_2$ & $3$ & $\left( 1,1 \right)_{X}$ \\  
\hline
\end{tabular}
\caption{Chiral spectrum of MSSM-like model.}
\label{mssmpec}
\end{table}

There are 7 $U(1)$s in the model but they all gain a Green-Schwarz mass. This can be checked by calculating the intersection of $\Pi_{U(1)}$, as in (\ref{piu1}), with the cycles $\{0,0,0,0,0\}_{+}$, $\{0,0,0,0,1\}_{+}$, $\{0,0,0,0,2\}_{+}$, $\{0,0,0,0,4\}_{+}$, $\{0,0,0,8,4\}_{+}$, $\{0,0,2,8,7\}_{+}$ and $\{0,0,2,8,8\}_{+}$ which give seven linearly independent conditions on the charges $Q_{a,b,c,d,e,f,g}$ and imply they vanish. The hypercharge
\be
U(1)_H = \frac16 U(1)_a + \frac12 U(1)_c + \frac12 U(1)_d + \frac12 U(1)_e \;,
\ee
satisfies a weaker version of (\ref{hyploc}) in that its intersection with the 5 visible branes vanish but its intersection with the 2 exotic ones does not. So that although it is not massless it is possible to add a brane wrapping $-\Pi_H$ so that a massless $U(1)$ arises with the correct charges to be hypercharge. The extra states that arise from this are labeled $\mathrm{F}_1$ and $\mathrm{F}_2$ in table \ref{pspec} and are due to non-vanishing intersections $I_{fH}=-I_{gH}=3$. Their charge under hypercharge is non-vanishing but undetermined.

\section{Discussion}

In this paper we studied model building using intersecting D6-branes on smooth CYs. We developed the techniques for dealing with a large class of CYs and studied some explicit examples producing Pati-Salam and MSSM-like models. These are the first chiral supersymmetric models constructed in this way.

There are a number of possible avenues for future study. It would be interesting to study how the singularity blow-up procedure used in constructing CYs within weighted projective spaces affects the constructions on special Lagrangians. If it is still possible to calculate intersection numbers using similar techniques to those used in this paper it would be possible to study model building on some of the other CYs constructed in \cite{Candelas:1989hd}. Some of these exhibit very large symmetry groups allowing for a large set of special Lagrangians which would improve the model building opportunities. Another way to enhance the class of available special Lagrangians is to study anti-holomorphic involutions associated with permutations.

One of the motivations for this work has been to improve the interactions between moduli stabilisation and model building. In particular the models of \cite{Palti:2008mg} developed a scenario with a dynamically low supersymmetry breaking scale but relied on CY compactifications and not torus orbifolds. It would be interesting to study scenarios with consistent chiral models and moduli stabilisation taking into account constraints such as those outlined in \cite{Blumenhagen:2007sm}. Finally it would be interesting to study how the non-perturbative instanton calculations in IIA initiated in \cite{Blumenhagen:2006xt} could be implemented within this model building framework.

I thank Volker Braun, Pablo Camara, Philip Candelas,  Joe Conlon, Rhys Davies, James Gray, Yang-Hui He, Dominic Joyce and Andre Lukas for useful and stimulating discussions. I especially thank Volker Braun and Joe Conlon for reading through the manuscript and for useful feedback. 

E.P. is supported by an STFC Postdoctoral Fellowship.
\newpage
\appendix

\section{Surface intersection numbers}
\label{sec:curandsur}

In this section we discuss intersections of special Lagrangians on loci of dimensions two (surfaces) and one (curves). The resulting intersection number is given by the self-intersection of the intersection locus. The self-intersection of a manifold can roughly be thought of as deforming the manifold along normal directions and counting its intersections with the undeformed version. The self-intersection number can be identified with the number of zeros that a section of the normal bundle must have. For special Lagrangian submanifolds the normal bundle is isomorphic to the tangent bundle\footnote{Given a basis $v_i$ with $i=1,2,3$ of the tangent bundle of the special Lagrangian, the basis one-forms $v_i \lrcorner J$ vanish when restricted to the special Lagrangian and form a basis for its normal bundle.}. For an intersection locus of special Lagrangian submanifolds the deformations must be normal to both the intersecting special Lagrangians and so the the common normal bundle is isomorphic to the tangent bundle of the intersection locus. Therefore the relevant intersection number is given by the number of zeros of sections of the intersection locus tangent bundle which is just its Euler character.  For curve intersections this always vanishes and so we are only interested in surface intersections.

To calculate the intersection number we therefore need to determine the topology of the intersection locus. For example for the quintic the surface intersection of the two special Lagrangians $\Pi_1 = \{ 0,0,0,0,0 \}_+$ and $\Pi_2 = \{ 0,0,0,0,1\}_+$ is given by 
\be
\xi_1^5 + \xi_2^5 + \xi_3^5 + \xi_4^5  = 0 \;,
\ee
with $\xi_i \in RP^3$. This has the topology of  $RP^2$ which can be seen by using the unique real solution for $\xi_1$ to map to an $RP^2$ spanned by $\{\xi_2,\xi_3,\xi_4\}$ \cite{Brunner:1999jq}. Therefore this surface intersections give an intersection number of $1$ which is the Euler character of $RP^2$. The sign of the intersection is just given by the relative orientation of the two special Lagrangians.

For other intersections we need to do a case-by-case determination of the topology. However, whenever the defining polynomial has a co-ordinate appearing with an odd power we can always repeat the quintic analysis and map surface intersections to $RP^2$ which has self-intersection $1$. Therefore the remaining cases are where all the powers in the polynomial are even. 
\newline
\newline
{\bf The cases $CP^4_{[5,2,1,1,1]}$ and $CP^4_{[2,1,1,1,1]}$:}
\newline
\newline
The analysis for these cases are essentially the same and so we discuss only $CP^4_{[2,1,1,1,1]}$. Our methodology is taken from \cite{Roiban:2002iv}. Consider surface intersections. There are two possibilities given by the polynomials
\ba
\mathrm{Case\;1:}\;\; \xi^6_2 + \xi_3^6 + \xi_4^6 - \xi_5^6 &=& 0 \;, \\
\mathrm{Case\;2:}\;\;\xi^6_2 + \xi_3^6 - \xi_4^6 - \xi_5^6 &=& 0 \;. \label{21111surint}
\ea
Consider case 1. The co-ordinates $\xi$ are projective but we can fix the rescaling freedom by replacing them with real affine co-ordinates, $\xi_{2,3,4}\in\mathbb{R}$  and imposing a homogeneity fixing constraint
\be
\xi_2^6 + \xi_3^6 + \xi_4^6 = 1 = \xi_5^6 \;.
\ee
This fixes the magnitude of the homogeneous rescaling parameter $\lambda$ but still leaves a $\mathbb{Z}_2$ redundancy associated with $\lambda=-1$. Therefore we need to mod out the topology by $\xi_i \rightarrow -\xi_i$. 
We therefore have two spheres at $\xi_5=\pm1$. The positive and negative configurations are related by the $\mathbb{Z}_2$. Therefore the topology is given by $S^2$ which gives an intersection number of $2$. For case 2 we can map this, by taking the third root and taking $\xi_{2,3,4,5} \in \mathbb{R}$ with
\be
\xi_2^2 + \xi_3^2 = 1 = \xi_4^2 + \xi_5^2 \;,
\ee
to $S^1 \times S^1$ (the $\mathbb{Z}_2$ just inverts the circle). This has self intersection 0.
\newline
\newline
{\bf The case $CP^4_{[4,1,1,1,1]}$:}
\newline
\newline
For surface intersections we have the cases
\ba
\mathrm{Case\;1:}\;\; \xi^8_2 + \xi_3^8 + \xi_4^8 - \xi_5^8 &=& 0 \;,\\ 
\mathrm{Case\;2:}\;\; \xi^8_2 + \xi_3^8 - \xi_4^8 - \xi_5^8 &=& 0 \;,\\ 
\mathrm{Case\;3:}\;\; \xi^2_2 + \xi_3^8 + \xi_4^8 - \xi_5^8 &=& 0 \;,\\
\mathrm{Case\;4:}\;\; \xi^2_2 + \xi_3^8 - \xi_4^8 - \xi_5^8 &=& 0 \;,\\
\mathrm{Case\;5:}\;\; \xi^2_2 - \xi_3^8 - \xi_4^8 - \xi_5^8 &=& 0 \;.
\ea
Cases 1 gives an $S^2$ which has self intersection $2$. Case 2 gives $S^1 \times S^1$ which has self intersection 0. Case 3 gives $S^2$ which has self intersection $2$. Case 4 gives an interval times a circle which has self intersection $0$. Case 5 gives $RP^2\cup RP^2$ which has self-intersection $2$.

\section{Example intersection calculation}
\label{sec:intexam}

In this section we present an example calculation of an intersection number between two special Lagrangians in the CY within $CP^4_{[5,2,1,1,1]}$ given by the vanishing polynomial
\be 
z_1^2 + z_2^5 + z_3^{10} + z_4^{10} + z_5^{10} = 0 \;.
\ee
The two cycles we consider are 
\ba
\Pi_{a} &=& \{ 0,0,0,7,3 \}_- \nn \\
\Pi_{b} &=& \{ 0,0,3,9,8 \}_+  \;.
\ea
The integers give the rotations of the co-ordinates in terms of powers of the 10th root of unity. So that for example $\Pi_a$ is given by the fixed point locus of the anti-holomorphic involution
\be
z_1 \rightarrow \bar{z}_1 \;, \;\;
z_2 \rightarrow \bar{z}_2 \;, \;\;
z_3 \rightarrow \bar{z}_3 \;, \;\;
z_4 \rightarrow e^{-\frac{2\pi i 7}{10}}\bar{z}_4 \;, \;\;
z_5 \rightarrow e^{-\frac{2\pi i 3}{10}}\bar{z}_5 \;.
\ee
We now set out to calculate their intersection. First we perform a co-ordinate change
\be
z'_4 = e^{\frac{\pi i 7}{10}} z_4 \;, \;\; z'_5 = e^{\frac{\pi i 3}{10}} z_5 \;,
\ee
so that the calculation becomes the intersection between 
\ba
\Pi_{0} &=& \{ 0,0,0,0,0 \}_+ \nn \\
\Pi_c &=& \{ 0,0,3,2,5 \}_-  \;,
\ea
and the defining polynomial becomes
\be 
z_1^2 + z_2^5 + z_3^{10} - \left(z'_4\right)^{10} - \left(z'_5\right)^{10} = 0 \;.
\ee
Note we have flipped both the orientations of the cycles so that the intersection number remains invariant.
We now go through the patches. For notation purposes we define 
\be
\alpha = e^{\frac{2\pi i}{10}} \;,
\ee
as in the main text. We also use $\xi_i$ and $x_i$ as defined in the main text.
\newline
\newline
{\bf Patch 1}
\newline
\newline
On this patch we have to count point solutions to 
\ba
x_1 &=& 1 \;, \nn \\
x_3 = x'_4 = x'_5 &=& 0 \;, \nn \\
1 + x_2^5 &=& 0 \;.
\ea
We have that $x_2 \in \mathbb{R} \times \left\{1,e^{\frac{2\pi i}{5}} \right\}$. There are two solutions at $(x_1,x_2)=\left\{(1,-1),(1,-e^{\frac{2\pi i}{5}})\right\}$ but they are related by a homogeneous transformation $\lambda = -e^{\frac{-\pi i}{5}}$ and so there is a single intersection. The sign of the intersection is negative since the cycles are of opposite orientation.

We now have to go through the rotations that are the orbifold symmetries of this patch. These are weighted rotations by fifth roots of unity that leave $\omega_1=1$. So for homogeneous parameter rotation $\lambda=\alpha$ we get
\be
\{ 0,0,3,2,5 \}_- \rightarrow \{ 0,4,5,4,7 \}_+ \;.
\ee
Note that the orientation has changed sign because a transformation, $\lambda = \alpha^n$ with $n$ odd, means that the calibration angle has increased by $2\pi$ as in (\ref{alor}). The resulting configuration has no solutions and so does not contribute an intersection number. Similarly we rotate
\be
\{ 0,4,5,4,7 \}_+  \rightarrow \{ 0,8,7,6,9 \}_- \rightarrow \{ 0,2,9,8,1 \}_+ \rightarrow \{ 0,6,1,0,3 \}_-\;\;.
\ee
Note that we have to keep track of orientation changes due to rotation angles going past the primary branch so that for example in the last rotation there are two minus signs coming from $9 \rightarrow 1$ and $8 \rightarrow 0$, and one minus sign from the calibration angle transformation giving an overall relative minus sign. The last configuration has two point solutions giving an intersection number of $-2$. So at the end of going through patch 1 the intersection number is $-3$.
\newline
\newline
{\bf Patch 2}
\newline
\newline
To study this patch we first rotate with $\lambda=\alpha^5$ so that 
\be
\{ 0,0,3,2,5 \}_- \rightarrow \{ 5,0,8,7,0 \}_+ \;.
\ee
On this patch we have $x_2=1$ but also only count solutions with $x_1=0$ in order to not over count. 

This configuration has two points that are solutions to the polynomial which are $x_1=x_3=x_4=0$ and $x'_5=\pm 1$. However the two points are identified by taking $\lambda=-1$ and so the configuration contributes an overall intersection number of $+1$.

There is a $Z_2$ orbifold rotation on this patch given by $\lambda=\alpha^5$ which gives
\be
\{ 5,0,8,7,0 \}_+ \rightarrow \{ 0,0,3,2,5 \}_- \;.
\ee
This has no solutions compatible with the constraint $x_1=0$ for patch 2 and so this configuration does not contribute any intersection numbers. Essentially we see that solutions here are just a repeat of the solutions found on patch 1.

So at the end of patch 2 the overall intersection number is $-2$. Note that we have a vector pair of intersections $-2 = -2 -1 + 1$. These are the Higgs vector pair of the Pati-Salam model of section \ref{sec:patisal}.
\newline
\newline
{\bf Patches 3, 4 and 5}
\newline
\newline
These patches do not have any orbifold symmetries associated to them and the configurations are given by
\be
\{ 0,0,3,2,5 \}_- \rightarrow \{ 5,4,0,9,2 \}_+ \rightarrow \{ 0,6,1,0,3 \}_- \rightarrow \{ 5,0,8,7,0 \}_+ \;.
\ee
None of these contribute intersection numbers since the intersections do not satisfy the constraints $x_1=x_2=0$. 

So the total intersection number is $-2$.

\section{Some models on $CP^4_{[2,1,1,1,1]}$ and $CP^4_{[4,1,1,1,1]}$}
\label{sec:someothermodels}

In this appendix we present a Pati-Salam-like model on the weighted projective space $CP^4_{[2,1,1,1,1]}$ and a two-generation $SU(5)$ GUT model on $CP^4_{[4,1,1,1,1]}$. These are simply to show that in general a weighted projective space is rich enough for a phenomenologically relevant model to be constructed.

A Pati-Salam-like model can be constructed on $CP^4_{[2,1,1,1,1]}$ as follows. The model is constructed from three stacks of branes $\{a,b,c\}$ with $N_a=4$, $N_b=2$ and $N_c=2$. The cycles that the branes wrap are
\ba
\Pi_a &=& \left\{0,0,0,1,5\right\}_{-} \;\;,\;\; \Pi_{a'} = \left\{0,0,0,5,1\right\}_{-} \;, \nn \\
\Pi_b &=& \left\{0,1,1,0,4\right\}_{-} \;\;,\;\; \Pi_{b'} = \left\{0,5,5,0,2\right\}_{+} \;, \nn \\
\Pi_c &=& \left\{0,1,1,3,1\right\}_{-} \;\;,\;\; \Pi_{c'} = \left\{0,5,5,3,5\right\}_{-} \;.
\ea
Here the integers stand for powers of the sixth root of unity. The intersection matrix for these cycles is given in table \ref{psint21111}.
\begin{table}
\center
\begin{tabular}{|c|ccccccc|}
\hline
\;& $\Pi_a$ & $\Pi_b$ & $\Pi_c$ & $\Pi_{a'}$ & $\Pi_{b'}$ & $\Pi_{c'}$ & $\Pi_0$ \\
\hline
$\Pi_a$ & $0$ & $-2$ & $2$ & $0$ & $-1$ & $1$ & $0$ \\  
$\Pi_b$ & \;  & $0$ & $0$ & \;  & $1$ & $-1$ & $1$\\  
$\Pi_c$ & \;  & \;  & $0$ & \;  & \;  & $1$ & $-1$ \\
\hline
\end{tabular}
\caption{Intersection numbers for Pati-Salam model on $CP^4_{[2,1,1,1,1]}$.}
\label{psint21111}
\end{table}
The intersection numbers give rise to the field spectrum given in table \ref{pspec21111}. 
\begin{table}
\center
\begin{tabular}{|c|c|c|}
\hline
Field & Multiplicity & Representation \\
\hline
$\mathrm{Q}_{\mathrm{L}}$ & $3$ & $\left( 4,2,1 \right)$ \\  
$\mathrm{Q}_{\mathrm{R}}$ & $3$ & $\left( \bar{4},1,2 \right)$ \\  
h & $1$ & $\left( 1,2,2 \right)$ \\  
\hline 
\hline 
$\mathrm{B}_{1}$ & $1$ & $[\mathrm{A}]_{SU(2)}$ \\  
$\mathrm{C}_{1}$ & $1$ & $[\mathrm{S}]_{SU(2)}$ \\  
\hline
\end{tabular}
\caption{Chiral spectrum of Pati-Salam model on $CP^4_{[2,1,1,1,1]}$. Note there is a missing heavy vector-like Higgs pair.}
\label{pspec21111}
\end{table}
The model lacks a heavy Higgs vector-like pair to complete to a full Pati-Salam model. Therefore this sector would have to arise from another sector in the string theory. The spectrum only has one charged chiral exotic denoted $C_1$. It is simple to check that the `local' tadpole conditions are satisfied and so a hidden sector added to cancel global tadpoles does not give rise to any new chiral matter charged under the visible gauge groups. 
All the $U(1)$s can be shown to gain a Green-Schwarz mass by intersecting the general $U(1)$ combination with the three cycles $\{0,0,0,0,0\}_{+}$, $\{0,0,0,1,0\}_{+}$, $\{0,0,1,0,2\}_{+}$ which imply $Q_a=Q_b=Q_c=0$.

A two-generation $SU(5)$ GUT model on $CP^4_{[4,1,1,1,1]}$ can be constructed as follows. The model is constructed from two stacks of branes $\{a,b\}$ with $N_a=5$, $N_b=1$. The cycles that the branes wrap are
\ba
\Pi_a &=& \left\{0,0,1,5,2\right\}_{-} \;\;,\;\; \Pi_{a'} = \left\{0,0,7,3,6\right\}_{+} \;, \nn \\
\Pi_b &=& \left\{0,1,0,1,6\right\}_{-} \;\;,\;\; \Pi_{b'} = \left\{0,7,0,7,2\right\}_{+} \;, 
\ea
Here the integers stand for powers of the eighth root of unity. The intersection matrix for these cycles is given in table \ref{psint41111}.
\begin{table}
\center
\begin{tabular}{|c|ccccc|}
\hline
\;& $\Pi_a$ & $\Pi_b$ & $\Pi_{a'}$ & $\Pi_{b'}$  & $\Pi_0$ \\
\hline
$\Pi_a$ & $0$ & $0$ & $-2$ & $2$ & $-2$  \\  
$\Pi_b$ & \;  & $0$ &  \;  & $-2$ & $2$ \\
\hline
\end{tabular}
\caption{Intersection numbers for two-generation $SU(5)$ GUT model on $CP^4_{[4,1,1,1,1]}$.}
\label{psint41111}
\end{table}
The intersection numbers give rise to the field spectrum given in table \ref{pspec41111}. 
\begin{table}
\center
\begin{tabular}{|c|c|c|}
\hline
Field & Multiplicity & Representation \\
\hline
$\mathrm{Q}$ & $2$ & $\bar{5}$ \\  
$\mathrm{E}$ & $2$ & $10$ \\  
\hline 
\hline 
$\mathrm{B}_{1}$ & $2$ & $1$ \\    
\hline
\end{tabular}
\caption{Chiral spectrum of two-generation $SU(5)$ GUT model on $CP^4_{[2,1,1,1,1]}$.}
\label{pspec41111}
\end{table}
It is simple to check that the `local' tadpole conditions are satisfied and so a hidden sector added to cancel global tadpoles does not give rise to any new chiral matter charged under the visible gauge group. 
The two $U(1)$s can be shown to gain a Green-Schwarz mass by intersecting the general $U(1)$ combination with the two cycles $\{0,0,0,0,0\}_{+}$, $\{0,0,0,1,7\}_{-}$ which imply $Q_a=Q_b=0$.



\begin{thebibliography}{99}

\bibitem{Berkooz:1996km}
  M.~Berkooz, M.~R.~Douglas and R.~G.~Leigh,
  ``Branes intersecting at angles,''
  Nucl.\ Phys.\  B {\bf 480}, 265 (1996)
  [arXiv:hep-th/9606139].

  R.~Blumenhagen, L.~Goerlich, B.~Kors and D.~Lust,
  ``Noncommutative compactifications of type I strings on tori with  magnetic
  background flux,''
  JHEP {\bf 0010} (2000) 006
  [arXiv:hep-th/0007024].

  G.~Aldazabal, S.~Franco, L.~E.~Ibanez, R.~Rabadan and A.~M.~Uranga,
  ``D = 4 chiral string compactifications from intersecting branes,''
  J.\ Math.\ Phys.\  {\bf 42}, 3103 (2001)
  [arXiv:hep-th/0011073].


\bibitem{Blumenhagen:2006ci}
  R.~Blumenhagen, B.~Kors, D.~Lust and S.~Stieberger,
  ``Four-dimensional String Compactifications with D-Branes, Orientifolds   and
  Fluxes,''
  Phys.\ Rept.\  {\bf 445} (2007) 1
  [arXiv:hep-th/0610327].

  F.~Marchesano,
  ``Progress in D-brane model building,''
  Fortsch.\ Phys.\  {\bf 55} (2007) 491
  [arXiv:hep-th/0702094].


\bibitem{Anderson:2008uw}
  L.~B.~Anderson, Y.~H.~He and A.~Lukas,
  ``Monad Bundles in Heterotic String Compactifications,''
  JHEP {\bf 0807} (2008) 104
  [arXiv:0805.2875 [hep-th]].

  M.~Gabella, Y.~H.~He and A.~Lukas,
  ``An Abundance of Heterotic Vacua,''
  JHEP {\bf 0812} (2008) 027
  [arXiv:0808.2142 [hep-th]].


\bibitem{Blumenhagen:2008zz}
  R.~Blumenhagen, V.~Braun, T.~W.~Grimm and T.~Weigand,
  ``GUTs in Type IIB Orientifold Compactifications,''
  arXiv:0811.2936 [hep-th].

\bibitem{Candelas:1987kf}
  P.~Candelas, A.~M.~Dale, C.~A.~Lutken and R.~Schimmrigk,
  ``Complete Intersection Calabi-Yau Manifolds,''
  Nucl.\ Phys.\  B {\bf 298}, 493 (1988).

\bibitem{Candelas:1989hd}
  P.~Candelas, M.~Lynker and R.~Schimmrigk,
  ``Calabi-Yau Manifolds in Weighted P(4),''
  Nucl.\ Phys.\  B {\bf 341}, 383 (1990).

\bibitem{Becker:1995kb}
  K.~Becker, M.~Becker and A.~Strominger,
  ``Five-Branes, Membranes And Nonperturbative String Theory,''
  Nucl.\ Phys.\  B {\bf 456}, 130 (1995)
  [arXiv:hep-th/9507158].

\bibitem{Brunner:1999jq}
  I.~Brunner, M.~R.~Douglas, A.~E.~Lawrence and C.~Romelsberger,
  ``D-branes on the quintic,''
  JHEP {\bf 0008}, 015 (2000)
  [arXiv:hep-th/9906200].

\bibitem{Blumenhagen:2002wn}
  R.~Blumenhagen, V.~Braun, B.~Kors and D.~Lust,
  ``Orientifolds of K3 and Calabi-Yau manifolds with intersecting D-branes,''
  JHEP {\bf 0207}, 026 (2002)
  [arXiv:hep-th/0206038].

\bibitem{Blumenhagen:2002vp}
  R.~Blumenhagen, V.~Braun, B.~Kors and D.~Lust,
  ``The standard model on the quintic,''
  arXiv:hep-th/0210083.

\bibitem{Uranga:2002pg}
  A.~M.~Uranga,
  ``Local models for intersecting brane worlds,''
  JHEP {\bf 0212} (2002) 058
  [arXiv:hep-th/0208014].

\bibitem{Beasley:2008dc}
  C.~Beasley, J.~J.~Heckman and C.~Vafa,
  ``GUTs and Exceptional Branes in F-theory - I,''
  JHEP {\bf 0901} (2009) 058
  [arXiv:0802.3391 [hep-th]].

  J.~P.~Conlon, A.~Maharana and F.~Quevedo,
  ``Towards Realistic String Vacua,''
  arXiv:0810.5660 [hep-th].

\bibitem{Palti:2008mg}
  E.~Palti, G.~Tasinato and J.~Ward,
  ``WEAKLY-coupled IIA Flux Compactifications,''
  JHEP {\bf 0806}, 084 (2008)
  [arXiv:0804.1248 [hep-th]].

\bibitem{Joyce:2001nm}
  D.~Joyce,
  ``Lectures on special Lagrangian geometry,''
  arXiv:math/0111111.

  D.~Joyce,
  ``Lectures on Calabi-Yau and special Lagrangian geometry,''
  arXiv:math/0108088.

\bibitem{paltifuture}
  E.~Palti,
  Work in progress...

\bibitem{Roiban:2002iv}
  R.~Roiban, C.~Romelsberger and J.~Walcher,
  ``Discrete torsion in singular G(2)-manifolds and real LG,''
  Adv.\ Theor.\ Math.\ Phys.\  {\bf 6}, 207 (2003)
  [arXiv:hep-th/0203272].

\bibitem{Hubsch:1992nu}
  T.~Hubsch,
  ``Calabi-Yau manifolds: A Bestiary for physicists,''
{\it  Singapore, Singapore: World Scientific (1992) 362 p}


\bibitem{Brunner:2003zm}
  I.~Brunner and K.~Hori,
  ``Orientifolds and mirror symmetry,''
  JHEP {\bf 0411}, 005 (2004)
  [arXiv:hep-th/0303135].

\bibitem{Klemm:1992tx}
  A.~Klemm and S.~Theisen,
  ``Considerations of one modulus Calabi-Yau compactifications: Picard-Fuchs
  equations, Kahler potentials and mirror maps,''
  Nucl.\ Phys.\  B {\bf 389}, 153 (1993)
  [arXiv:hep-th/9205041].

\bibitem{Blumenhagen:2007sm}
  R.~Blumenhagen, S.~Moster and E.~Plauschinn,
  ``Moduli Stabilisation versus Chirality for MSSM like Type IIB
  Orientifolds,''
  JHEP {\bf 0801} (2008) 058
  [arXiv:0711.3389 [hep-th]].

\bibitem{Blumenhagen:2006xt}
  R.~Blumenhagen, M.~Cvetic and T.~Weigand,
  ``Spacetime instanton corrections in 4D string vacua - the seesaw mechanism
  for D-brane models,''
  Nucl.\ Phys.\  B {\bf 771} (2007) 113
  [arXiv:hep-th/0609191].

  L.~E.~Ibanez and A.~M.~Uranga,
  ``Neutrino Majorana masses from string theory instanton effects,''
  JHEP {\bf 0703} (2007) 052
  [arXiv:hep-th/0609213].





\end{thebibliography}
\end{document}